*Review*

# Survey of Context Information Fusion for Sensor Networks based Ubiquitous Systems

**Vijay Borges[1*] and Wilson Jeberson[2]**

[1]  Department of Computer Sc. & I.T., SHIATS-DU, Allahabad, India; E-Mail:
    vb@gec.ac.in (Vijay Borges)
[2]  Department of Computer Sc. & I.T., SHIATS-DU, Allahabad, India; E-Mail:
    jeberson@rediffmail.com (Wilson Jeberson)

*  Author to whom correspondence should be addressed; E-Mail: vb@gec.ac.in (Vijay
   Borges);
   Tel.: +91-992-379-9058; Fax: -

**Abstract**: Sensor Networks produce a large amount of data. According to the needs this data requires to be processed, delivered and accessed. This processed data when made available with the physical device location, user preferences, time constraints; generically called as context-awareness; is widely referred to as the core function for ubiquitous systems. To our best knowledge there is lack of analysis of context information fusion for ubiquitous sensor networks. Adopting appropriate information fusion techniques can help in screening noisy measurements, control data in the network and take necessary inferences that can help in contextual computing. In this paper we try and explore different context information fusion techniques by comparing a large number of solutions, their methods, architectures and models.

## 1. Introduction

*''The most profound technologies are those that disappear. They weave themselves into the fabric of everyday life until they are indistinguishable from it.''* So began Mark Weiser's seminal 1991 paper [1] that described his vision of ubiquitous computing, now also called pervasive computing. The essence of that vision was the creation of environments saturated with devices with computing and communication capability, yet gracefully integrated with human users. This vision is slowly seeing the days of realization, through the rapid development of the Wireless Sensor Networks deployments in many areas of our lives.

A Wireless Sensor Network (WSN) is a kind of an ad hoc network consisting of a large number of nodes fitted with different sensor devices [2]. The objective of WSN may be to gather data, monitor an event etc so that necessary actions could be taken as required. WSN generates a large amount of data; so the basic need is to process this large collected data. In addition to that the data generated may be noisy, redundant and intermittent due to the failures of the underlying sensor nodes[2]. Information fusion arises as a means to how this gathered data can be processed to increase the relevance from the data collection. As humans will be more and more involved in this pervasive environment; generating context information to supplement human efforts would be an added advantage. The ability to recognise what a user is doing or the situation how a group of users are involved in task collaborations could be activities where pervasive applications reaction, adaptation and aid in future activities would be highly desirable. Pervasive applications could span from health-care monitoring to smart home and office automation, from intelligent sightseeing guides to new generation gaming.

Given the importance of context information fusion in an ubiquitous environment based on WSN's, this survey highlight the niche areas related to context information fusion and how it has been used in an ubiquitous way for sensor based systems. To achieve context information fusion in a least intrusive way requires an integrated sensor based ubiquitous systems. This is challenging since sensor based systems are highly heterogeneous, have severe communicating and computing constraints, and operating in challenging environments. Context information fusion works across protocol layers (physical layer up to application layer), this adds to the challenge of designing a uniform model.

In this survey the background on context information fusion would be presented. Various classification methods would be discussed next. Latest architectures would then be discussed along with its pros and cons. Finally concluding what kind of research efforts have gone in the area of context information fusion.

## 2. Fundamentals

Mark Weiser in his seminal paper defined a vision called '*Ambient Intelligence'* (1) where many different devices will gather and process information from many sources to both control physical processes and interact with human beings. These technologies should be unobtrusive (ubiquitous). One of the critical aspects required is to transfer relevant information (context) to the place where it is needed. To bring this envisioned technology into the fore wireless communication is critical. Therefore a class of networks called Wireless Sensor Network (WSN) [2] came into being to fill the gap. These networks consist of individual nodes that are able to interact with their environment by sensing or controlling physical parameters; these nodes collaborate with other nodes to complete their tasks. These tasks could be event detection, periodic measurements, tracking etc. Apart from the tasks which the WSN could achieve there are certain characteristics [2] desired of WSN like; Type of Service, Quality of Service, Fault tolerance, Lifetime, Scalability, Range of Density, Programmability, Maintainability.

## 2.1. WSN Architecture and Constraints

A WSN consist of a collection of sensor nodes. These nodes comprise five main components: Controller, Memory, Sensor and actuator, Communication and Power Supply. Each of these components operates balancing between minimising energy consumption and fulfilling assigned tasks.

### 2.1.1 Controller

The controller is the core of the wireless sensor node. It collects data from the sensors, processes this data, decides when and where to send it, receives data from other sensor nodes, and decides on the actuator's behaviour. It has to execute various programs, ranging from time-critical signal processing and communication protocols to application programs. This controller can be a Microcontroller, Microprocessor, Field-Programmable Gate Array (FPGA) or Application Specific Integrated Circuit (ASIC).

### 2.1.2 Memory

The memory stores intermediate sensor readings, packets from other nodes, programs modules to achieve tasks. The memory component could be variants of Random Access Memory (RAM) and Read Only Memory (ROM). RAM stores the information till power supply is available, while ROM retains its contents past the power supply shutdown. Generally access time from RAM is faster than ROM. Variants of ROM's which allows data to be re-written could be Electrically Erasable Programmable Read-Only Memory (EEPROM) or flash memory.

### 2.1.3 Sensor and Actuator

Sensor is a device that detects a change in a physical stimulus in the environment and turns into a signal which can be measured or recorded. The stimulus can be acoustic, electric, magnetic, optic, thermal, mechanical etc. [2]. An actuator is a mechanism by which a control system acts upon an environment. It works by converting energy into motion. Actuators can be hydraulic, pneumatic, electric, mechanical etc. [3].

### 2.1.4 Communication

Turning nodes into a network requires a device for sending and receiving information over a wireless channel. Generally for wireless communication Radio Frequency (RF) based communication is the best choice due to long range, high data rates, acceptable error rates at low energy consumption, and no requirement for line-of-sight between sender and receivers.

### 2.1.5 Power Supply

Generally for no tethered power supply batteries provide energy to the sensor nodes. Alternatively recharging can be obtained from the environment (e.g. solar, ambient noise)

### 2.2. Ubiquitous Computing Environment

In his seminal paper Mark Weiser popularised the term '*Ubiquitous Computing*' [1]. Ubiquitous computing (also called pervasive computing) is an environment which is saturated with objects having computing and communicating capabilities. According to [4], pervasive computing incorporates four thrust areas. '*Effective use of smart environments*'; by incorporating embedded computing infrastructure in a building infrastructure, creates a smart space that brings these two worlds together [5]. The second thrust is '*invisibility*'; is the complete disappearance of pervasive computing technology from the user's consciousness. The thrust research area is '*localized scalability*'; as smart spaces grow in sophistication, the intensity of interactions between a user's personal computing space and his surrounding increases. These interactions place severe demands on bandwidth, and energy of the embedded infrastructure. The last thrust is '*masking uneven conditioning*' of environment; which handles on issues of masking the truly smart spaces from dumb spaces due to economic reasons.

### 2.3. Context Aware Computing

Context awareness as an essential ingredient of ubiquitous and pervasive computing systems existed from the early 1990s. Mark Weiser coined '*ubiquitous computing*' and [6] came with '*context-aware*'. "*Context is any information that can be used to characterize the situation of an entity. An entity is a person, place, or object that is considered relevant to the interaction between a user and an application, including the user and the applications themselves*"[7]. [7] Goes on to define '*Context-awareness*' as, "*A system is context-aware if it uses context to provide relevant information and/or services to the user, where relevancy depends on the user's task*". Thus context type can be categorised as present activity, identity, location, and time. The categorisation of context awareness can be presentation of information service to a user, automatic execution of a service, and tagging of context for later retrieval. '*Context-aware Computing*' is a style of computing in which situational and environmental information about people, places, and  things is used to anticipate needs and proactively offer enriched, situation-aware and usable content, functions, and experiences.

## 3. CONTEXT INFORMATION DISTRIBUTION

WSN is very prone to node failures, yet it is very robust and fault tolerant. To overcome sensor failures, technological limitations, spatial, and temporal coverage problems, certain properties must be ensured: cooperation, redundancy, and complementarity [7][8]. In WSN deployment scenarios a region of interest is covered using many nodes, each cooperating with a partial view of the scene; context information fusion can be used to compose the complete view by piecing together from each nodes. Redundancy makes WSN almost transparent of single a node failure; overlapping measurements can be fused for more accurate data [9]. Complementarity is achieved using sensors that perceive different properties of the environment; context information fusion can be used to combine complementary context information so that it allows inferences that may otherwise have been difficult to obtain from individual node measurements.

### 3.1. Context and QoC Definition

Many authors address context. In [10], service context is addressed as, *"where you are, who are you with, and what resources are nearby"*; [7]refers to it as, *"information that can be used to characterize the situation of an entity"*; [11] categorizes it as: individual activity, location, time, and relations; [12] refers to context as *"set of variables that may be of interest for an agent and that influences its actions"*; [13] divides context into four-dimensional space, computing context, physical context, time context and user context.

[13] Refers to Computing context, to encapsulate all technical aspects related to computing capabilities and resources. This encapsulation is necessary as it expresses all the heterogeneities present in the mobile environment; like device capabilities and connectivity.

The physical context arranges into groups, aspects from the real world that are accessible by sensors/actuators deployed in the surrounding. Aspects such as traffic conditions, speed, noise levels, temperature and lighting data are addressed [14]. Problem with physical context are measurement errors due to imprecision of the physical processes.

Time context addresses the time dimension, such as time of day, week, month and season of the year, of the activity performed by the system. These activities could be sporadic events, whose occurrences are triggered occasionally; or periodic events that occur in a predictable and repeatable way [7].

Finally, user context contain high-level context aspects related to the social dimension of users (got from users being part of a whole system), such as user's profile, people nearby, and current social situation [15].

Quality of Context (QoC), refers to the set of parameters that express quality requirements and properties for context data (precision, freshness, trustworthiness) [16][17]. [18]deals with context data with four QoC parameters (i) being up-to-date to deal with data aging; (ii) trustworthiness to the rate the belief we have in the context correctness; (iii) completeness to consider that context data could be partial and incorrect; (iv) significance to express differentiated priorities; (v) context data validity, specifies validity to be complied by the context data; and (vi) context data precision, evaluates degree of adherence between real, sensed and distributed value of context data. QoS does not require perfect context data but rather a correct estimate of the data quality.

### 3.2. Context Information Distribution in Ubiquitous Environment

Context-aware services should only have to produce and publish context information and declare their interests in receiving, and must also handle issues with context information distribution. Context information distribution deals with automatically delivering of this context information to all entities who have expressed interest in it. There can be two types of context distribution. Uniformed context information distribution, which simply routes context data according to context needs expressed by nodes (publish/subscribe systems). Nodes routes the context information without examining the content. The other type is the informed context information distribution, wherein the exchanged context information is dynamically adapted and self-managed to assist the distribution process itself.

### 3.3. Necessities for Context Information Distribution

There has been a steady rise in the way context-aware distribution is done. Earlier the research focus was on small scale deployments like smart home or smaller infrastructure deployments. Currently the changes are to adapt the wireless context-aware deployments in large scale deployments often reaching the internet scales. To support such large context-aware deployments there are many shortfalls that require to be fulfilled: (a) Context information distribution to route produced information to all interesting sinks in the system; (b) Support for heterogeneous sensor nodes with varied capabilities ranging from computing speeds, communicating standards, different operational scenario etc.; (c) Presenting varied visibility scopes for context information, taking into consideration physical locality, user reference context; so as to limit management overheads; (d) QoC-based constraints fulfilments like, quality of the received information, adaptation based on the topology changes, meeting delivery guarantees, timeliness and reliability and avoiding redundant and conflicting copies in the system; (e) End-to-end Context-information life cycle management [19]. Activities like distributed information aggregation and filtering have to be handled to reduce unnecessary management overheads.

### 3.3. Context Information Distribution

The context information distribution logical architecture as adapted from [20] is as shown in Figure 1. This architecture envisions three principal actors: context source, context sink and context distribution function. Context source masks back-end sensors' access operations and enables context data publishing. Context sink permits the service level to express its context needs by either context queries (pull-based interactions) or subscriptions (push-based interactions); context matching is the correct satisfaction of the sink requests. Context distribution entity distributes context by mediating the interaction between context sources and sink, by automatically notifying subscribed context sinks on context matching. There are other supporting entities in the architecture Context Management, Context Delivery and Runtime Adaptation Support.

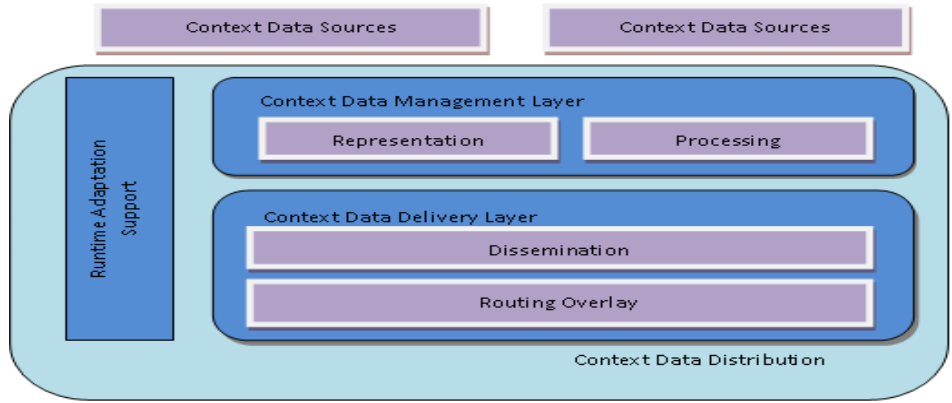

Figure 1. System architecture of a context distribution system

### 3.3.1 Context Management Entity

Context Management entity would be responsible for the local context handling by defining context representation and expressing processing needs and operations. Context representation includes different models and techniques as shown in Figure 2. These models could be classified according to [21][22] as General Model, Domain-specific models, No Model. They could be so classified to differ in expressiveness, memorisation costs and processing overheads. General model offers generic problem representation of the knowledge. Domain-specific models, represents only data belonging to specific domain and avoiding generic representation of knowledge. No model, do not focus on knowledge representations. Generic models have different formalism and expressiveness and have adapted the widely accepted models like: key-value model, markup scheme models, logic-based models, and ontology-based models [23][22].

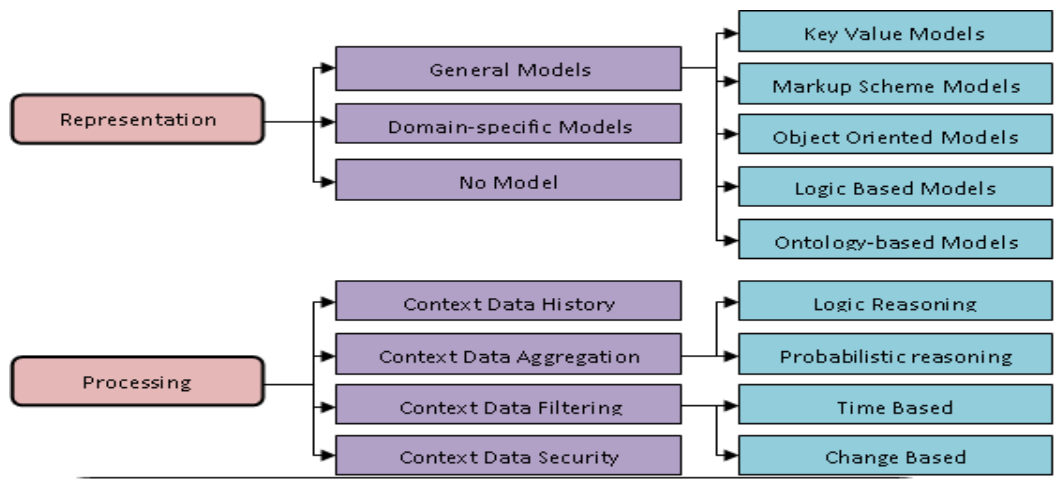

Figure 2. Classification of the Information Context Management Entity

Key-value models, represents the simplest data structure for modelling context by exploiting pairs of two items: key (attribute name) and its value. It is simple for implementation and thus is popular. It has its own failings, since it lacks capabilities for structuring context data and has no means for checking data validity. Context Toolkit, work from [24] adopts this approach to represent both context and metadata associated with context sources. Pervasive Autonomic Context-aware Environment (PACE) [25] depends on key-value pairs to represent context data used to determine which action the user prefers in the current ubiquitous context. History-Based routing protocol for Opportunistic networks (HiBOp) and Context-aware Adaptive Routing (CAR), use computing, time and user context to evaluate and select the best forwarder.

Markup scheme models use XML-based representations to model hierarchical data structure consisting of markup tags, attributes and contents. They are advantageous over key-value pairs like, (i) validating context data via XMl-schemas, (ii) structuring data via XML structures. Context-aware Resource Management Environment (CARMEN) exploits XML-based profiles to describe both computing and user context information [26]. Context Casting (C-CAST) uses context provisioning aspects and defines an XML-based Context Meta Language (ContextML) to distribute context data into the system [27]. Context Sharing In uNreliable Environments (COSINE) builds a modular context sharing in which contexts are represented by XML and can be queried by using XPath queries [28].

Object-oriented models, take advantages of the features of the object-oriented paradigm especially encapsulation and reusability. Each class defines a new context type with access functionality, type-checking and data validity at runtime and compile time; QoC parameters can be easily mapped in objects. Use of object abstractions simplifies the deployment of context handling code. Context entities composition and Sharing (COSMOS), each context is exemplar as an object comprehending several built-in mechanisms to ensure push- and pull-based change notifications [29]. ReconFigureurable Context-Sensitive Middleware (RCSM) uses an Interface Definition Language (IDL); by using it the developer can specify context/situations relevant to the application, the actions to trigger and the timings of these actions [30].

Logic-based models, take advantage of the high expressiveness intrinsic to the logic formalism: context contains facts, expressions and rules, while new knowledge can be delivered by inference. These models have limitations on the validity of the context. [31][32] discuss using first order predicate logic to represent context as a quaternary predicate **(<ContextType>, <Subject>, <Relater>, <Object>)**; where **<ContextType>** is the context type the predicate is describing; **<Subject>** is the person, place, or physical object the context is concerned; **<Object>** is the value associated with the **<Subject>**; and **<Relater>** links **<Subject>** and **<Object>** by means of a comparison operator **(=,>,<)**, a verb, or a preposition. CORTEX and Context-awareness Sub-Structure (CASS) fall in this category [33][34].

Ontology-based models, use ontology's to represent context. This focus on relationships between entities, as ontology's are apt at mapping everyday knowledge within a data structure, reuse of previous works and creation of common and shared domain vocabularies. Service-Oriented Context-Aware Middleware (SOCAM) composes generic as well as domain specific ontology's [35]. SOCAM classifies data as direct – sensed by sensors or defined by users –and indirect – derived by inference. Context Broker Architecture-OWL (CoBrA-Ont) uses context knowledge base and OWL-based ontology to memorize available knowledge [36]. Ontology-models and Logic-based models are generally avoided in the Sensor network scenarios due to the resource-constraints of the sensor nodes.

Spatial models are used widely for localization systems to represent real-world objects' locations. *MiddleWhere* is location-aware based context distribution system [37].

Context processing which is the other half in the Context Management entity, includes both (i) production of new knowledge from pre-existing context by using aggregation techniques (matching, first-order logic aggregation, semantic-based etc.); and (ii) simple filtering techniques to aid system scalability, by context distribution to currently available resources, [21]. Security of the context also plays a important part in context processing.

Context aggregation techniques are based on logic and probability reasoning, based on whether the system considers the context correct or correct to a certain degree. Aggregation techniques though resource crunchy are nonetheless fundamental to enable context-awareness since, (i) difficulty in defining context due to huge amount of possible context directions, and (ii) context undergoes continuous updates which has to be done automatically. Logic- or Ontology-based models are the two directions apt for dynamic data aggregation.

### 3.3.2 Context Delivery Entity

Context Delivery Entity would be responsible for routing the context into the ubiquitous system. This entity would generally be above the network infrastructure. It has got two core components, dissemination and routing overlay, depicted in Figure 3. Dissemination deals with; (i) which context to have distributed; and (ii) which destination nodes will receive the distributed data. Routing overlay, considers that context distribution could exploit different overlay networks to connect and organize the involved brokers.

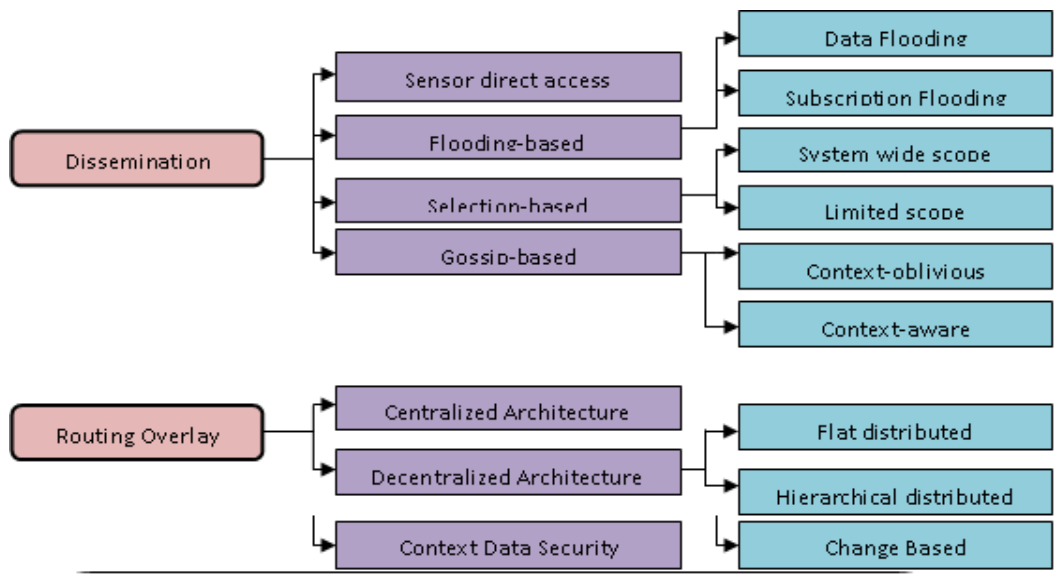

Figure 3. Classification of the Information Context Distribution Entity

The dissemination module enables context flow between sources and sinks. Dissemination solutions are; sensor direct access, flooding-based, selection-based, and gossip-based. In sensor direct access sinks communicate directly with sources to access data. Context Toolkit [24] discoverers handle registration from context sources and enable device mobility. COSMOS [29] focuses on context processing assuming all the context data are produced by local sensors. RCSM [30] implements a context discovery protocol to manage registrations of local sensors and discover remote sensors, on application start up. In flooding-based algorithms context dissemination is achieved via flooding operations of the context or of the subscription. In context flooding, each node broadcasts known context to spread them in the system by letting receiver nodes locally select context to receive. In case of Adaptive Traffic Lights exchanges context useful for coordinating red/yellow/green times between vehicles near an intersection [38]. Selection-based algorithms have two parts. First it deterministically builds dissemination backbones by using context subscriptions; in the next step dissemination happens only between these backbones and only interested nodes. Visibility of the entire system or a limited scope (set of nodes) can be achieved. Gossip-based algorithms disseminate data in a probabilistic manner letting each node resend the context to a randomly-selected set of neighbours. They are well suited for fast-changing and instable networks like the WSN. There is a variant called the context-aware gossip-based protocol, which is typically used for selecting neighbours for gossiping based on context belonging to very different context dimensions. These membership criteria's could be social similarity [39], distance between nodes [40] etc.

Routing overlay takes care of organising the brokers involved in context dissemination. Architecturally it could be centralized or decentralized. Centralized architectures includes a possible concentrated deployment; while decentralized could be flat or hierarchical distribution.

### 3.3.3 Runtime Adaptation Support

Runtime adaptation support deals with dynamically managing and modifying context data distribution (Figure 4). Classification of the runtime adaptation according to [20] could be; (i) unaware, (ii) partially-aware, and (iii) totally-aware. In unaware adaptation, the service level neither reaches nor influences runtime adaptation. In partially-aware adaptation, there is more collaboration between the service level which supplies profiles that describe the required kind of services requests and the runtime adaptation which modifies context data distribution to meet those requests. In totally-aware adaptation, the runtime adaptation support does not perform anything on its own, but it is the service level that completes drives reconfigurations.

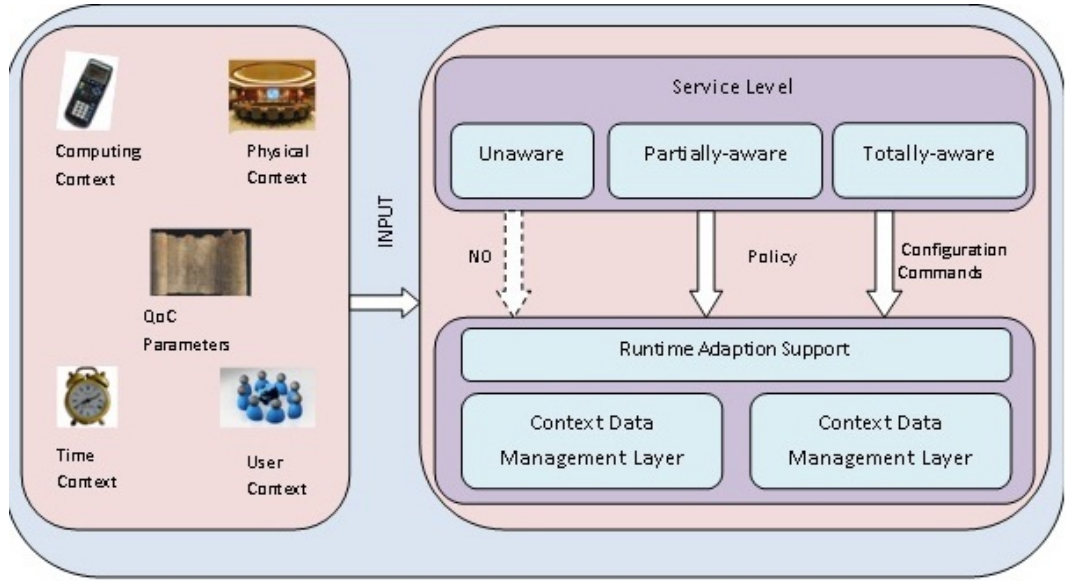

Figure 4. Runtime adaptation support

## 4. Classification of Context Information Fusion

WSN was designed primarily to gather and process data from the environment in order to have a better understanding of the behaviour of the monitored entity [2]. This generated data is more useful if the context related to the production of this data is captured. Context information fusion concerns with how this contextual information gathered by sensors can be processed to increase its relevance. Contextual information fusion can be commonly used in detection and classification tasks, such as robotics and military applications [41], intrusion detection [42] and Denial of Service (DoS) detection [43].

Context information fusion can be categorized into three categories according to [7]; (i) based on relationships among input context; (ii) based on abstraction level of the manipulated context during fusion process; and (iii) based on the abstraction level of the input and output of a fusion process.

Context information fusion based on relationship between the input contexts can be further classified as complementary, redundant, or cooperative [44]. In complementary context information fusion, when context information is provided by different sources, context information fusion obtains a piece of context information that is more complete. An example of complementary context information fusion that fuses information from sensor nodes into a feature map that describes the whole sensor field is dealt in [45][46][47]. In redundant context information fusion, if two or more independent sources provide the same piece of context information, these pieces can be fused to increase the associated confidence [7]. In cooperative context information fusion, two independent sources cooperate when the context information provided by them is fused into new context information, which is more informative [41].

Context information fusion based on levels of abstraction is sub-classified into low-level fusion, medium level fusion, high-level fusion, or multilevel fusion [7]. In low-level fusion (signal/measurement level fusion) as dealt in [48] is achieved by applying moving average filter to estimate ambient noise to infer availability of the communication channel. In medium-level fusion (feature/attribute level fusion) [49][47], attributes or features of an entity (shape, texture, position) are fused to obtain feature map. In high-level fusion (symbolic/decision level fusion), symbolic representation are taken as combined inputs to obtain higher level of confidence or achieve a global decision. [50] Uses Bayesan approach for binary event detection as an example of higher-level fusion. In multi-level fusion both the input and output of fusion can be of any level. [51] Uses Dempster-Shafer theory as an example of multi-level fusion to decide node failures based on traffic decay features.

Context information fusion based on abstraction level of the input and output is further sub-divided according to [52] into five categories. Data In-Data Out (DAI-DAO), this fusion deals with raw data and the result is also more reliable/accurate raw data. Data In-Feature Out (DAI-FEO), uses raw data from sources to extract features or attributes that describe an entity. Feature In-Feature out (FEI-FEO), works on a set of features to improve/refine a feature, or extract new ones. Feature In-Decision Out (FEI-DEO), takes a set of features of an entity generating a symbolic representation or a decision. Decision In-Decision Out (DEI-DEO), decision is fused in order to obtain a new decision.

### 4.1. Mechanisms and Algorithms for Context Information Fusion

Context information fusion can be performed with different objectives such as inference, estimation, classification, feature maps, and compression.

Inference methods are generally applied in decision context fusion, where decision is taken based on perceived situational knowledge. Classical methods are based on Bayesian inference and Dempster-Shafer Belief Accumulation theory. Context information fusion based on Bayesian inference offers formalism to combine evidence based on rules of probability theory. Bayesian inference is based on Bayes' rule [53]: $Pr(A|B)=Pr(B|A)Pr(A)/Pr(B)$; where the posterior probability $Pr(A|B)$ states the belief in the hypothesis A given the information B; the probability $Pr(A)$ is the prior probability and the probability $Pr(B)$ is treated as the normalising constant. The criticality in Bayesian formalism is that $Pr(B|A)$ and $Pr(A)$ have to be estimated or guessed apriori. [54] Uses neural network to estimate the conditional probabilities to feed the Bayesian inference module for decision-making. [50] Uses this method for event detection in WSN. The infer algorithm of [55] uses this method to determine missing data from the nodes that are not active. The other classical work on inference is the Dempster-Shafer Inference (Theory of Evidence) [56][57] that generalizes the Bayesian theory. It uses beliefs or mass functions, like Bayes' rule uses probabilities. It can be used even when there is incomplete knowledge representation, belief updates, and evidence combination [58]. A key concept in Dempster-Shafer reasoning system is the 'frame of discernment', which is a set of all possible states that describe the system and the states are exhaustive and mutually exclusive. The elements of the power set of these states are called hypothesis. A probability is assigned to every hypothesis; based on probability theory Dempster-Shafer defines the belief function 'bel' and degree of doubt 'dou' on the hypothesis. Dempster-Shafer theory allows for information fusion of sensory contexts [59], and it allows source to contribute information with different levels of details, without need to assign apriori probabilities to unknown propositions (which can be later assigned when supporting information is available). In [60], the Data Service Middleware (DSWare) for WSN uses this theory assign a confidence value to every decision. In [51] uses this theory to improve the tree-based routing algorithms by detecting routing failures, and triggering a route re-discovery when absolutely needed. Others techniques of Inference methods are Fuzzy Logic, Neural Networks, Abductive Reasoning and Sematic Information Fusion. Fuzzy logic approximates reasoning to draw (possibly imprecise) conclusions from imprecise premises. [61] Uses intelligent sensor network and fuzzy logic control for autonomous navigational robotic vehicle that avoids obstacles. Neural Networks [62], uses input/output pairs as examples to generalize and build supervised learning mechanisms. Kohonen maps are examples of unsupervised neural networks [63]. Generally neural networks can be used in learning systems with fuzzy logic used to control its learning rate [64][65]. [66] Uses a fusion scheme to create edge maps of multi-spectral sensor images from radars, optical sensors, and infrared sensors. In Abductive reasoning, a hypothesis is chosen that best explains observed evidence [67]. Semantic Information fusion is done as in-network inference process on raw sensor data. It has two phases: knowledge base construction and pattern matching (inference). The first phase aggregates the most appropriate knowledge abstractions into semantic information, which is used in the second phase for pattern matching, for fusing relevant attributes and providing semantic interpretation of sensor context information.

Estimation methods are incorporated from control theory and use probability theory to compute a process state vector from a (or sequence) measurement vector [68]. Some of the methods used here are Maximum Likelihood, Maximum A Posteriori, Least Squares, Moving Averages filter, Kalman filter, and Particle filter. In Maximum Likelihood, wanting to compute the context of information fusion state 's', and having a set 'z' = $\{z(1), z(2), .., z(k)\}$ of k observations of 's'; the likelihood function $\lambda(s) = pdf(z|s)$ {pdf: probability density function}. The Maximum Likelihood estimator (MLE) looks out for the value of 's' that maximizes the likelihood function $x\hat{}(k) = \arg\max_x pdf(z|s)$. MLE is used to solve discovery problems; to obtain accurate distance estimations [69][70][71]. Maximum A Posteriori (MAP) is based on Bayesian theory, when a parameter x to be discovered is based on the outcome of a random variable with known pdf p(s). Given a set 'z' = $\{z(1), z(2), .., z(k)\}$ of k observations of 's'; the MAP estimator searches for the value of s that maximises the posterior distribution function $x\hat{}(k) = \arg\max_x pdf(s|z)$. Least Squares method is an optimization technique that searches for a function that best fits a set of input measurements. This is achieved by minimizing the sum of the square error between the points generated by the function and the input measurement. This method does not assume any prior probability, hence it works in a deterministic manner. This method quickly converges but is effected by noisy measurements. [47] Uses this method in guiding mobile nodes to build spatial maps. The Moving Average filter [72] is adopted in digital signal processing, as it reduces random white noise while retaining sharp step response. Thus is used in processing encoded signals in the time domain. Kalman filter [73] is used to fuse low-level redundant data. A issues in using Kalman filters in WSN is that it requires clock synchronisation among sensor nodes.

Feature Maps methods are used in applications such as guidance and resource management. In applications where raw sensory data is difficult to use, features representing aspects of the environment can be extracted and used by the requesting application using methods of estimation and inference. There two major types of feature maps: occupancy maps and network scans. Occupancy maps define a 2D/3D representation of the environment, describing which areas are occupied by an object and which areas are free. The observed space is divided into square cells containing values that indicate its probability of being occupied. Network Scans defined in [45] is a sort of resource/activity map for WSN. These maps indicate the distribution of the resources or activity of a WSN.

Compression methods employed in WSN exploit spatial correlation among sensor nodes with no extra communication cost. This is done by observing that two neighbours provide correlated measurements. In Distributed Source Coding (DCS) [74] data is compression from sources that are physically separate, and not communicating. The sources send their compressed output to central unit for joint decoding. In another method called Coding by Ordering [75], every node in a region of interest sends its data to a border node, which is responsible for grouping all packets into a super-packet which is then sent to the sink node. The important property that is extracted here is that border nodes can suppress some packets and sort the remainder (when order is not important), such that the values of the suppressed packets can be automatically inferred. In [76], presents a simple algorithm using energy efficient lossless compression technique based on Huffman coding scheme, where it exploits the natural correlation between the data and principles of entropy. The runtime of this algorithm shows it is much more efficient that other compression tools like gzip, bzip2, and S-LZW [77][78][79].

## 4.2. Context Information Fusion Architectural Models and Deployments

Several architectures and models serve as guidelines to design the context information fusion systems. Following architectural models that are apt to be applied context information fusion context in ubiquitous environment would be touched in this sub-section: Information-based model, activity-based model, and role-based model. A complete discussion on the others models for generic Wireless Sensor Networks are dealt in [80]. The context information-based model focuses on the abstraction level of the information handled by the fusion tasks. These models do not specify the execution sequence of the fusion tasks. In the context activity-based models, the activities and their correct sequence of execution are explicitly specified. In context Role-based models information fusion systems can be modelled and designed based on the fusion roles and the relationships among them. They however do not specify fusion tasks, instead provide a set of roles and specify the relationships among them.

Architectures based on context information-based systems are centred on the abstraction of the data generated during context fusion. The JDL model [81] and the Dasarathy model [52] are two variants in this class. The JDL model was conceived jointly by the U.S. Joint Directors of Laboratory (JDL) and U.S. Department of Defense (DOD).

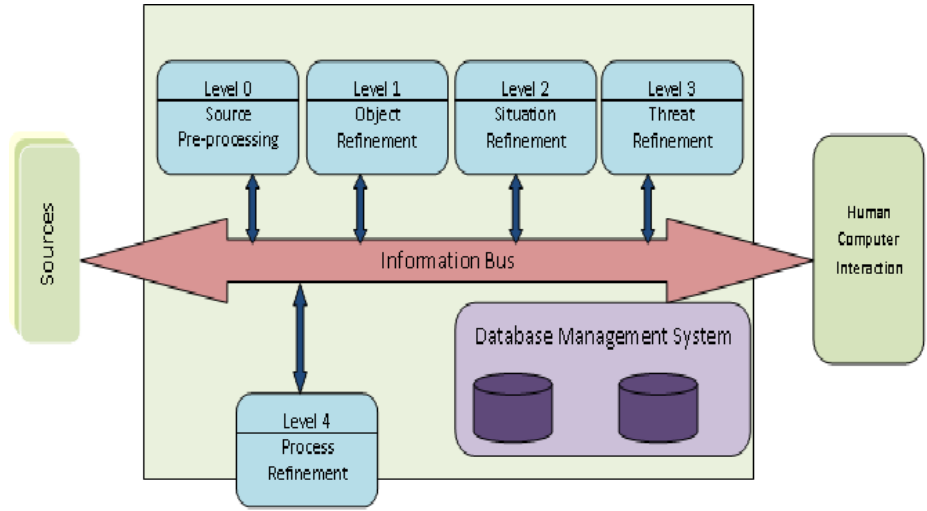

Figure 5. The JDL Model

As depicted in the Figure 5, JDL has five processing levels, an associated database, and an information bus connecting all components. Sources provide the input information fed from the sensors, human interface, databases etc. The Database Management System, handles the critical function of dealing with large and varied amount of data. This system can be adapted to handle the context coming in from the WSN deployments in the environment, and can handle queries efficiently without interacting with the individual context deployments. The Human Computer Interaction (HCI), allows human inputs, commands, queries, notification fusion of alarms, displays, graphics, and sounds. Level 0 (Source Preprocessing) aims at allocating context information to appropriate processes and selecting appropriate sources. Level 1 (Object Refinement), transform the context information into a consistent structure. Level 2 (Situation Refinement), provides a contextual description of the relationship among objects and observed events. Level 3 (Threat Refinement), evaluates the current context projecting it into the future to identify possible threats. Level 4 (Process Refinement), is a meta-process responsible for monitoring the system performance and allocating the sources based on set goals.

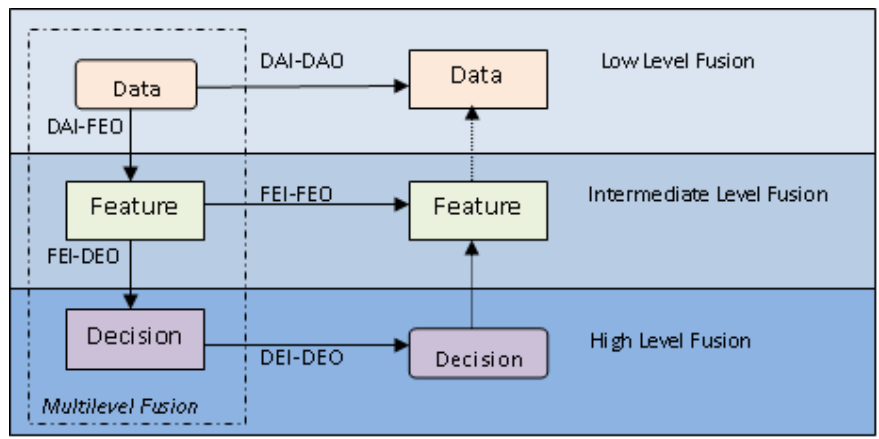

Figure 6. The DFD model

The Dasarathy Model or the DFD (Data-Feature-Decision) is depicted in Figure 6, is a context information fusion model based on inputs and outputs. The primary input is raw data and the main output is a decision. DFD model is used as ambient noise estimation [48], feature map building [47], event detection [82], and failure detection [51].

Architectures based on context activity-based models are based on the activities that must be performed in their correct sequence of execution. The Omnibus Model [83] organises the stages of context information fusion system in a cyclic sequence, based on the Observe-Orient-Decide-Act (OODA) loop [84]. It deals with the context gathering from the WSN deployment.

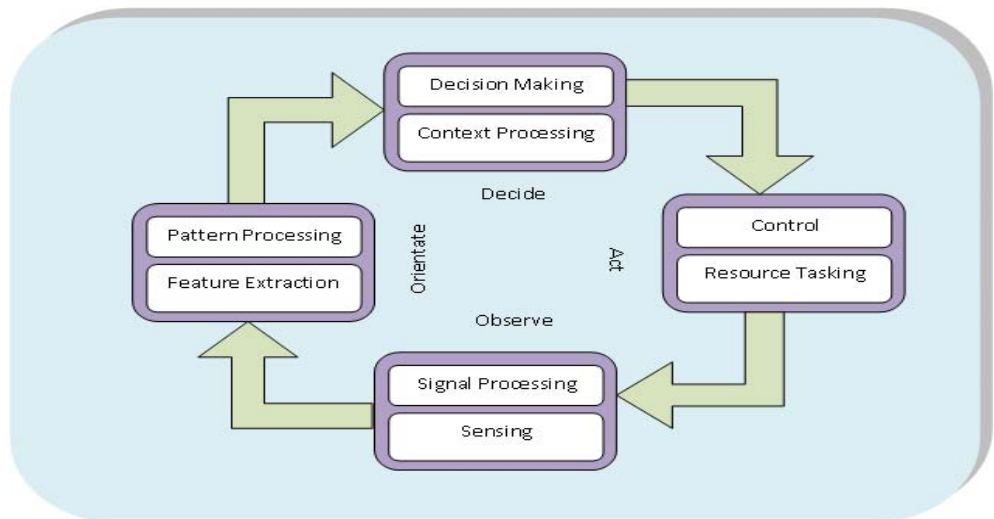

Figure 7: Omnibus Model

As depicted in the Figure 7, the first step in the Omnibus Model, Sensing and Signal Processing stage (Observe), information is gathered and pre-processed. In the Feature Extraction stage (Orient), from the gathered information, patterns are extracted and generally fused to create necessary contexts. The Decision stage the context is processed and actions to be followed are laid down. Similarly if there are threats in the system can be trapped in this stage. In the Act stage, the laid down action plans are acted upon by choosing the best plan to follow.

Architectures based on context role-based model can be best exemplified by focussing on the Object-Oriented Model [85]. The object-Oriented Model shown in Figure 8, uses cyclic architecture. There are however no fusion tasks or activities. The roles identified are Actor, Perceiver, Director, and Manager. The Actor is based with the interaction with the world, collecting information and acting on the environment. The Perceiver assesses the information and provides contextualized analysis to the director. The Director comes with an action plan taking into consideration the system's goals. Finally, the Manager controls the actors to execute the plans as stipulated by the director.

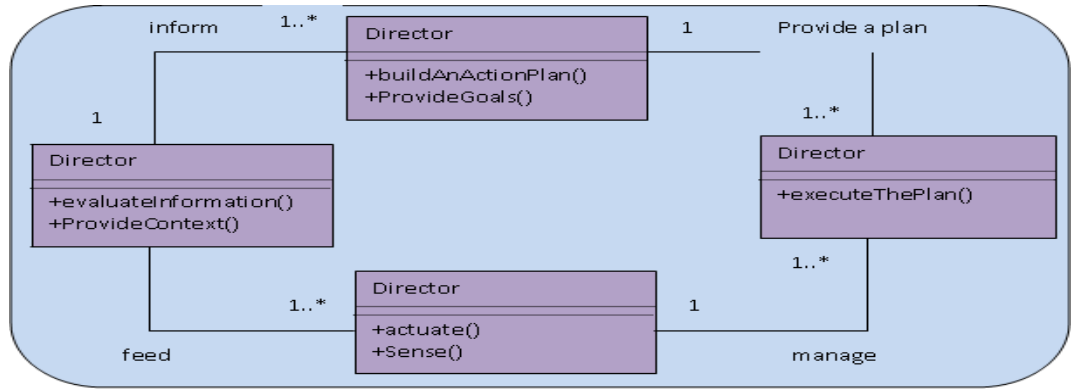

Figure 8. The Object-Oriented model for Context information fusion

## 5. Context Information Fusion Frameworks

Context information fusion frameworks should be able to understand the available context source (physical and virtual), their data structure, and automatically built internal data model's to facility them. The raw context needs to be retrieved and transformed appropriately into context representation models with negligible human aid. The frameworks must be flexible to support multi-modal reasoning, while having access to contextual information both real-time as well as historic. Frameworks to support Context-as-a-Service (CXaaS) has been discussed in [86], the life cycle is classified into Enterprise Lifecycle Approaches (ELA) and Context Lifecycle Approaches (CLA). ELA concentrates on context whereas CLA dwells into context management. ELA circle around *'information lifecycle'* (creating, receipt, distribution, use, maintenance, and disposition); *'enterprise content management'*; '*Observe, Orient, Decide, Act'* OODA/Boyd loop [84]. CLA lifecycles works around context sensing, context transmission, context acquisition, context classification, context handling, context dissemination, context usage, context deletion, context maintenance, context disposition [86].

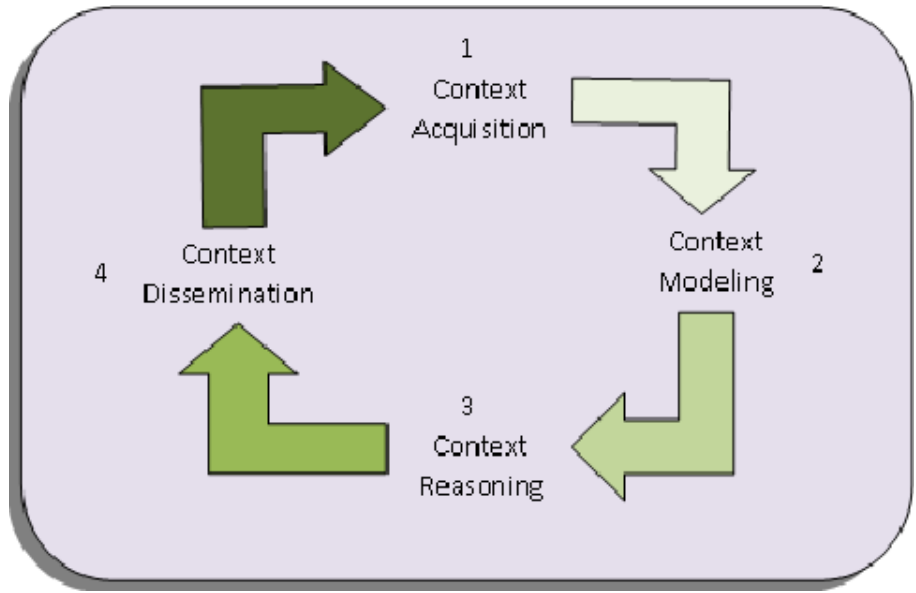

Figure 9. The Context Life Cycle

The simplest context life cycle can be put in four phases as shown in the Figure 9 [87]. In the context acquisition phase gets the needed context from various relevant sources. The techniques to acquire context is based on responsibility, frequency (context sent periodically or on exceeding a threshold limit), context source (sensor hardware, middleware, context servers), sensor type (physical, virtual, logical sensors), and acquisition process (directly from sensor, infer for sensor data, provided manually). The collected contexts are modelled and represented according to a meaningful schema. The modelling can be Key-Value model, markup model, graphical model (Unified Modelling Language, Object Role Model), logic based model, and ontology based model. The modelled contexts are processed to derive high-level (reasoning) context information. The context reasoning can be defined as a method of deducting new knowledge, based on the available context. Context reasoning has three important phases. Context pre-processing deals with context cleaning (fill missing values, handle outliers, validate context via multiple sources, etc.), context fusion, and context inference. Context reasoning is categorised as: supervised learning, unsupervised learning, rules, fuzzy logic, ontological reasoning, and probabilistic reasoning. This context information is then distributed to consumers who have registered using query, subscription method. In the query method of context distribution context consumers make a request in terms of a query, which gets processed by the context management systems to produce results. In the subscription (publish/subscribe) method, the system returns the result periodically or when the event occurs.

Context reasoning techniques can be computationally intensive and time consuming when the context data is large [88][89][90]. Ontologies, are the preferred mechanism of managing and modelling context, which are based on semantic techniques [91]. [92] Defines ontology as: "*Ontology is a formal, explicit specification of a shared conceptualisation. A conceptualisation refers to an abstract model of some phenomenon in the world by having identified the relevant concepts of that phenomenon. Explicit* *means that the type of concepts used, and the constraints on their use are explicitly defined. For example, in medical domains, the concepts are diseases and symptoms, the relations between them are casual and a constraint is that a disease cannot cause itself. Formal refers to the fact that the ontology should be machine readable, which excludes natural language. Shared reflects the notion that an ontology captures consensual knowledge, that is, it is not private to some individual, but accepted by a group".* There are two steps in developing ontology's. First, the domain and scope need to be clearly defined. The existing ontology's are reviewed to find possibilities of leverage in existing ones. This is the main goal of ontology's; reusability of shared knowledge, interoperability among context-aware systems, and support for inference/reasoning. The growing interest in the adaptation of ontology's and ontological reasoning to automatically recognize complex context data resulted in the emergence of Web Ontology Language (OWL). OWL has developed into a standard for semantic web, and is supported by a number of tools for knowledge engineering and reasoning. [93] Shows solutions using extensive experimental evaluation and simulations for different intelligent environments using specifically OWL2. OWL2 language constructs are apt for activity representation; its axioms can be used to represent certain rules and rule-based reasoning in hybrid approaches having unique semantics by avoiding inconsistencies.

## 6. Research Efforts in Context Information Fusion

A complete survey of the research efforts in context aware computing can be found in[87]. Some key toolkits / middleware's are show in this survey with concentration on context information fusion.

**Context Toolkit** [24] aims to facilitating development and deployment of context-aware applications. It has three main abstraction: context widget (to retrieve data from sensors), context interpreter (reasoning about sensor data), and context aggregator.

**CoBrA** [94] (Context Broker Architecture) is a broker centric agent architecture that provides knowledge sharing and context reasoning for smart spaces. It mainly addresses supporting resource-limited mobile computing devices and privacy issues. Context information is modelled using ontology's, and it uses context brokers. A context broker has four components: context knowledge base (persistent storage for context in formation), context reasoning engine (reasoning over context information stored), context acquisition module (retrieve context from context sources), and policy management module (manages policies, such as who has access to what data). Context knowledge is represented in Resource Description Framework (RDF) triples using Jena.

**SOCAM** [35] (Service Oriented Context-Aware Middleware) is an ontology based context-aware middleware. Ontology's is separated into two levels: upper level ontology for general concepts and lower level ontology's domain specific descriptions. It has the following key components: context provider (acquires data from sensors and other internal and external data sources and converts the context in to Web Ontology Language (OWL) representation [95], context interpreter (performs reasoning using reasoning engine and stores processed context information in the knowledge base), context-aware services (context consumers), and services locating service (context providers and interpreter are allowed to register so other components can search for appropriates providers and interpreters based on their capabilities).

**e-SENSE** [96] combines body sensor network (BSN), object sensor networks (OSN), and environment sensor network (ESN) to capture context-rich information. It stages are sensor data capturing, data pre-filtering, context abstraction data source integration, context extraction, rule engine, and adaption.

**MoCA** [97] is a service based distribution middleware that employs ontology's to model and manage context. The context management nodes (CMN), is infrastructure that is responsible for managing the context domain. The main components in MoCA are: context providers (generating or retrieving context from other sources available to be used by the context management system), context consumers (consume the context gathered and processed by the system), and context service (responsible for receiving, storing, and disseminating context information). It uses an object oriented model for context handling. Extendible Markup Language (XML) is used to model context and check validity. The program codes acquire context and insert the data into context repository.

**Feel@Home** [98] is a context management framework that supports interaction between different domains. It is demonstrated in smart home, smart office, and mobile domains. The context information is stored using Web Ontology Language (OWL). It has three parts: user queries, global administration server (GAS), and domain context manager (DCM). User queries are first received by GAS. It decides what the relevant domain needs to be contacted to answer to answer the user query. Then GAS redirects the user query to the relevant domain context managers. DCM consists of typical context management components such as context wrapper, context aggregator, context reasoning, knowledge base, and several other components to mange user queries, publish/subscribe mechanism. The answers to the user query will return by using the same path as when received.

**ezContext** [99] is a framework that provides automatic context life cycle management. ezContext comprises several components: context source (physical sensors, databases or webservice), context provider (retrieves context from various sources whether in push/pull method), context manager (handles context modelling), context wrapper (encapsulates retrieved context into correct format), and providers' registry (list context providers and their capabilities). JavaBeans are used as the main data format.

CAMPUS [100] is a middleware exploiting technologies from semantic computing to dynamically derive adaptation decisions according to run-time contextual information. It is based on three essential technologies: compositional adaptation, ontology, and description logic/first-order logic reasoning; to construct context-aware adaptation decisions. It frees developers from the need to predict, formulate, and maintain adaptation rules, thereby greatly reducing the efforts required to develop context-aware applications.

## 7. Conclusion

Over the last few years improvement in sensor hardware technology at reduced costs would result in their attachment to the objects around us truly creating a ubiquitous environment around us keeping true to the vision of Mark Wieser. The main challenge lies in understanding the enormous contextual information that would be generated by these sensor deployments. This challenge is being taken up actively by public/private corporate as well as research institutes. In this survey paper, the effort has been to analyse and evaluate the information context fusion research efforts. Analysis has been made on a number of models, architectures and solutions that have cropped up from the research efforts of hundreds of individuals and multitudes of research institutes. The outcome of this survey strongly points to the research direction the community is taking in regards to information context fusion. This paper tries to give some ground work by analysing the efforts in this area as done in the past so that futuristic efforts would be more fruit bearing. The trend shows that this area is an active hub and much more efforts are required to have a truly ubiquitous environment around us.

## Acknowledgments

The authors would like to thank Almighty God, our families and Prof R.B. Lal, Prof Newman Fernandes, SHIATS, Allahabad; Prof Ramanujam, IMSc, Chennai; Prof D. Manjunath, IIT, Bombay; for their timely help and guidance. Work on this project is funded by the research grant through Computer Society of India vide their grant no 1-14/2013-09 dated 21/03/2013.

## Conflicts of Interest

The authors declare no conflict of interest.